# A Dualheap Selection Algorithm – A Call for Analysis

Greg Sepesi
sepesi@eduneer.com
http://www.eduneer.com

**ABSTRACT**

An algorithm is presented that efficiently solves the selection problem**:** finding the k-th smallest member of a set. Relevant to a divide-and-conquer strategy, the algorithm also partitions a set into small and large valued subsets. Applied recursively, this partitioning results in a sorted set. The algorithm's applicability is therefore much broader than just the selection problem.

The presented algorithm is based upon R.W. Floyd's 1964 algorithm that constructs a heap from the bottom-up. Empirically, the presented algorithm's performance appears competitive with the popular quickselect algorithm, a variant of C.A.R. Hoare's 1962 quicksort algorithm. Furthermore, constructing a heap from the bottom-up is an inherently parallel process (processors can work independently and simultaneously on subheap construction), suggesting a performance advantage with parallel implementations.

Given the presented algorithm's broad applicability, simplicity, serial performance, and parallel nature, further study is warranted. Specifically, worst-case analysis is an important but still unsolved problem.

**keywords:** algorithm, dualheap, parallel, partition, selection, sorting

## 1. ALGORITHM

A heap is an array with elements regarded as nodes in a complete binary tree, where node j is the parent of nodes 2j and 2j+1, and where the value at each parent node is superior to the values at its children's nodes. This superiority is commonly called the heap condition.

The dualheap selection algorithm is based upon an algorithm, first suggested in 1964 by R.W.Floyd [1], that establishes the heap condition in a bottom-up fashion. Floyd's heap construction algorithm is employed to create two heaps. Then a variant of Floyd's heap construction algorithm is employed to exchange node values until all of the nodes in the first heap have values larger than any node in the second.

Figure 1-1 lists an implementation of Floyd's heap construction algorithm. As noted in the listing, it is easy to change the sense of the heap so the smallest value instead of the largest is at the heap's root. The listing of Figure 1-2 is an implementation that uses negative indices to change the heap's direction, resulting in a heap with its root at the array's last element instead of the first.

Symbolically, the upward pointing triangle of Figure 1-3 represents the large valued heap constructed by the algorithm of Figure 1-1, and the downward pointing triangle represents the small valued heap constructed by the algorithm of Figure 1-2. I refer to this configuration of two opposing heaps as a dualheap and I use an arrow next to the dualheap to show the direction of increasing values.

The area of heap overlap in Figure 1-3 represents the nodes containing values that must be exchanged in order for all nodes in the large valued (bottom) heap to have values larger than any node in the small valued (top) heap. I refer to the portion of the dualheap selection algorithm that exchanges node values between heaps as the swapping phase. The implementation shown in the listing of Figure 1-4 is a variation of Floyd's heap construction algorithm. Instead of enforcing the heap condition at all nodes, it enforces the heap condition for only those nodes visited during a greedy traversal. This traversal visits nodes with the large values in the small valued heap and the small values in the large valued heap. During the traversal, node values are exchanged between heaps and the heap condition is maintained in a bottom-up fashion.

Finally, Figure 1-5 lists an implementation of the dualheap selection algorithm, showing the repeated use of Floyd's heap construction algorithm. Although presented as a sequential algorithm, one of the dualheap selection algorithm's most promising features is its parallel nature. For instance to construct a heap of size $n = 2^m - 1$ when there are $p = 2^q$ processors available, each of the p processors could independently and simultaneously construct a subheap of size $2^{(m-q)} - 1$. The remaining $2^q - 1$ downheap operations necessary to complete the heap construction could also be distributed among the p processors. (Note: $n = 2^m - 1 = 2^q (2^{(m-q)} - 1) + 2^q - 1$)

```
/* some shared variables (to reduce            /* some shared variables (to reduce
** argument passing)                           ** argument passing)
*/                                             */
static int *pl; /* large heap base */          static int *ps; /* small heap base */
static int lhn; /* # heap elements */          static int shn; /* # heap elements */

/***********************************           /***********************************
** DownLarge : sink kth element to             ** DownSmall : sink kth element to
** proper position in heap                     ** proper position in heap
**                                             **
** NOTE:                                       ** NOTE:
**  '>' puts larger values near root           **  '>' puts larger values near root
**  '<' puts smaller values near root          **  '<' puts smaller values near root
*/                                             */
void DownLarge(int k) {                        void DownSmall(int k) {
   int j, v;                                      int j, v;

   v = pl[k];                                     v = ps[-k];
   j = 2*k;                                       j = 2*k;
   j += (pl[j+1]<pl[j])?1:0; /*note*/             j += (ps[-j-1]>ps[-j])?1:0; /*note*/
   if (pl[j]<v) {           /*note*/              if (ps[-j]>v) {            /*note*/
      do {                                           do {
         pl[k] = pl[j];                                 ps[-k] = ps[-j];
         k = j;                                         k = j;
         j = 2*k;                                       j = 2*k;
         if (j > lhn) break;                            if (j > shn) break;
         j += (pl[j+1]<pl[j])?1:0;                      j += (ps[-j-1]>ps[-j])?1:0;
                           /*note*/                                       /*note*/
      } while (pl[j]<v);    /*note*/                 } while (ps[-j]>v);    /*note*/
      pl[k] = v;                                     ps[-k] = v;
   }                                              }
}                                              }

void ConstructLargeHeap() {                    void ConstructLargeHeap() {
   int i;                                         int i;
   for (i=lhn/2; i>0; i--)                        for (i=shn/2; i>0; i--)
      DownLarge(i);                                  DownSmall(i);
}                                              }
```

**Figure 1-1. Construction of Large Valued Heap**   **Figure 1-2. Construction of Small Valued Heap**

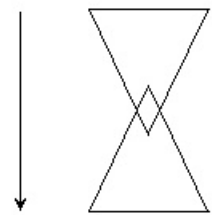

**Figure 1-3. Dualheap**

```
/*********************************
** TreeSwap
**
** ARGUMENTS:
**     (int) element of small heap
**     (int) element of large heap
*/
void TreeSwap(int ks, int kl) {
   int js, jl, tmp;

   js = 2*ks;
   jl = 2*kl;
   if ((js<=shn) && (jl<=lhn)) {
      js += (ps[-js-1]>ps[-js])?1:0;
      jl += (pl[jl+1]<pl[jl])?1:0;
      if (ps[-js] > pl[jl]) {
         TreeSwap(js,jl);
         if (ps[-(js^1)] > pl[jl^1])
            TreeSwap(js^1,jl^1);
      }
   }
   tmp = ps[-ks];
   ps[-ks] = pl[kl];
   pl[kl] = tmp;

   if (ks <= shn/2) DownSmall(ks);
   if (kl <= lhn/2) DownLarge(kl);
}
```

**Figure 1-4.  Greedy Traversal/Swap**

```
/*****************************
** Select
**
** ARGUMENTS:
**     (int *) array
**     (int) # elements
**     (int) selection index
**
** RETURN VALUE:
**     (int) value @ selected index
**
** NOTES:
** 1) array's 1st element is at
** index 1,
** 2) the number of elements
** does not include the data
** sentinels (see note 3),
** 3) at index 0 store a value
** guaranteed to be as low as any
** value in the array, and at
** index n+1 store a value as
** large as any value in the
** array.
*/
void Select(int *ph, int n, int k)
{
   int i;

   /*********************
   ** construction phase
   */

   /* Construct heap. */
   lhn = n;
   pl = ph;
   for (i=lhn/2; i>0; i--)
      DownLarge(i);

   /* Construct the small heap. */
   shn = ((k&0x1)==0)?k-1:k;
   ps = ph + shn + 1;
   for (i=shn/2; i>0; i--)
      DownSmall(i);

   /* Construct the large heap. */
   lhn = n - shn;
   pl = ph + shn;
   for (i=lhn/2; i>0; i--)
      DownLarge(i);

   /*********************
   ** swapping phase
   */
   while (ps[-1]>pl[1])
      TreeSwap(1,1);

   return ph[k];
}
```

**Figure 1-5.  Dualheap Selection Algorithm**

Note that even though dualheap select returns the k-th smallest value, an important side effect of the algorithm is that all elements at indices less than k will have values at least as small and all elements at indices greater than k will have values at least as large.

## 2.  ALGORITHM DEVIATIONS

The swapping phase's TreeSwap, that exchanges trees (2-dimensional), could be replaced by the simpler BranchSwap implemented in Figure 2-1 that exchanges branches (1-dimensional) or replaced by the even simpler RootSwap implemented in Figure 2-2 that exchanges just the roots (0-dimensional). However Figures 2-3 and 2-4 show the performance of TreeSwap, BranchSwap, and RootSwap in terms of the number of node value comparisons and node value moves, and TreeSwap performs the best.  With TreeSwap, the performance of the swapping phase is apparently **O**(n) although a proof of this remains an open problem.  RootSwap performs the worst, which is not surprising given swapping with RootSwap is basically an implementation of HeapSort [2] which has **O**(n*log(n)) performance [3].

```
/*******************************
** BranchSwap
*/
void BranchSwap() {
   int ks, kl;
   int js, jl, tmp;

   /* Traverse greedy path. */
   for (ks=1,kl=1,js=2,jl=2;
      (js<=shn) && (jl<=lhn);
      ks=js,kl=jl,js*=2,jl*=2)
   {
      js+=(ps[-js-1]>ps[-js])?1:0;
      jl+=(pl[jl+1]<pl[jl])?1:0;
      if (ps[-js] <= pl[jl])
         break;
   }

   /* Exchange values and
   ** restore heap condition,
   ** in bottom-up fashion.
   /
   for (; kl>=1; ks/=2, kl/=2) {
      tmp = ps[-ks];
      ps[-ks] = pl[kl];
      pl[kl] = tmp;

      if (ks <= shn/2)
         DownSmall(ks);
      if (kl <= lhn/2)
         DownLarge(kl);
   }
}
```

**Figure 2-1.  BranchSwap**

```
/*******************************
** RootSwap
/
void RootSwap() {
   int tmp;

   tmp = ps[-1];
   ps[-1] = pl[1];
   pl[1] = tmp;

   DownSmall(1);
   DownLarge(1);
}
```

**Figure 2-2.  RootSwap**

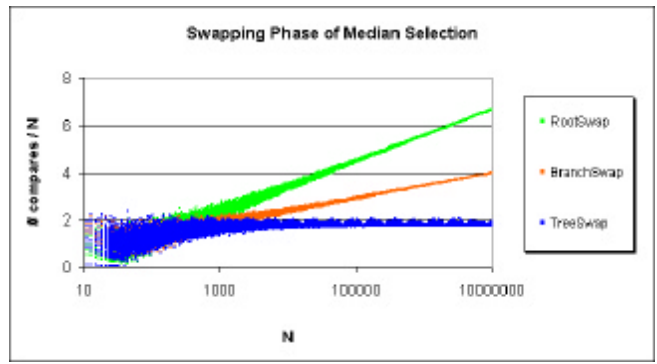

**Figure 2-3.  # Compares by Swapping Phase**

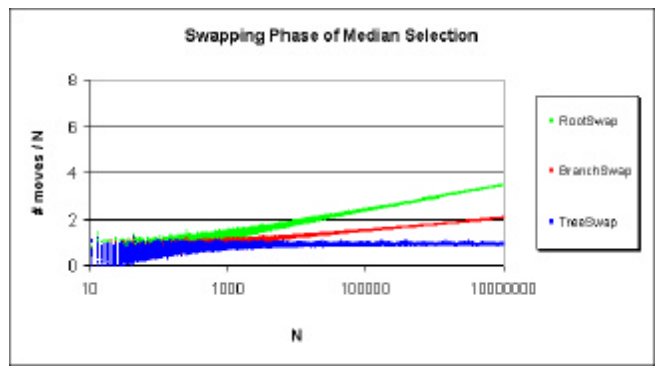

**Figure 2-4.  # Moves by Swapping Phase**

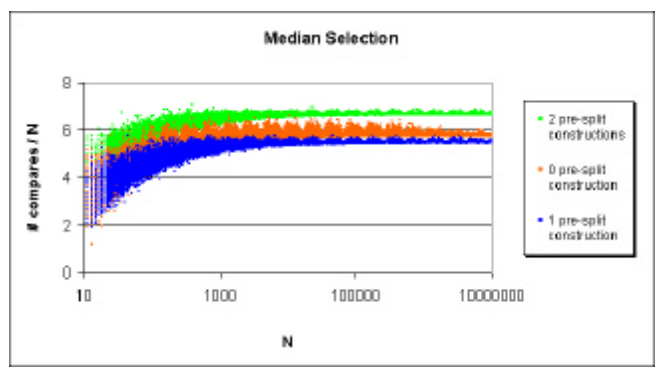

**Figure 2-5.  # Compares by Both Phases**

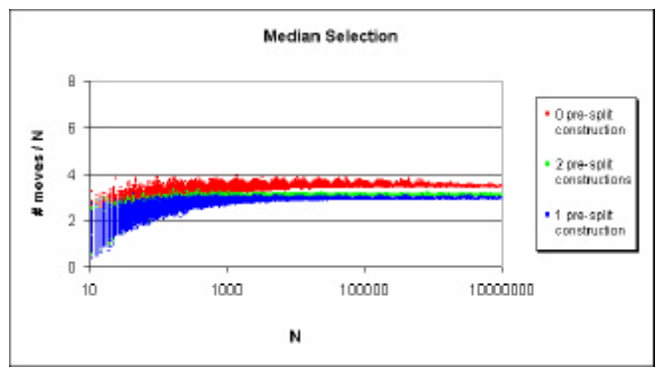

**Figure 2-5.  # Moves by Both Phases**

Having justified the complexity of the swapping phase, justifying the construction phase, specifically the pre-split heap construction, is next. Referring to Figure 1-5, I call the initial heap construction a pre-split construction because it occurs before the split into large valued and small valued heaps. An obvious question is, "What is the purpose of constructing a heap, if the heap is immediately split and reconstructed into small and large valued heaps?" It turns out that the seemingly unnecessary pre-split heap construction results in less dualheap overlap and, somewhat surprisingly, an improved overall algorithm performance. Figures 2-5 and 2-6 show the overall (both construction and swapping phases) performance of dualheap select with no pre-split heap construction, with one pre-split heap construction (as listed in Figure 1-5), and with two pre-split heap constructions (a small valued heap construction inserted after the large valued heap construction but before the split and reconstruction into large and small valued heaps). The best performance is with one pre-split heap construction, as listed in Figure 1-5. (Note that complete source code and test data for this report's figures are available at http://www.eduneer.com/)

## 3. ALGORITHM COMPARISONS

The performance of the popular quickselect algorithm, a variant of C.A.R.Hoare's quicksort algorithm [4], depends upon its choices of pivot values. A pivot value is the value at the junction of the small and large subsets about which elements get exchanged. Values smaller than the pivot value end up in the small valued subset and values larger than pivot value end up the large valued subset. Since good choices of pivot values are near the median value, and poor choices of pivot values result in quadratic worst-case performance [5], a median-of-medians algorithm [6] is often employed as a median estimator. The median-of-medians algorithm guarantees a linear worst-case performance, but it causes a significantly slower average-case performance.

Pivoting about values near the median is also important to the performance of dualheap select. However, unlike quickselect that requires a fairly complex estimator to determine a good pivot value, dualheap select pivots about addresses instead of values and can simply calculate the median address (halfway through the input array). Figures 3-1 and 3-2 show the number of value comparisons and the number of value moves, respectively, for quickselect, quickselect with a median estimator, and dualheap select. Quickselect has a better average-case performance than dualheap select and it appears that dualheap select has a better average-case performance than quickselect with a median estimator, and a better worst-case performance than both quickselect and quickselect with a median estimator. However, worst-case analysis of dualheap select is necessary, but presently missing, to verify these observations.

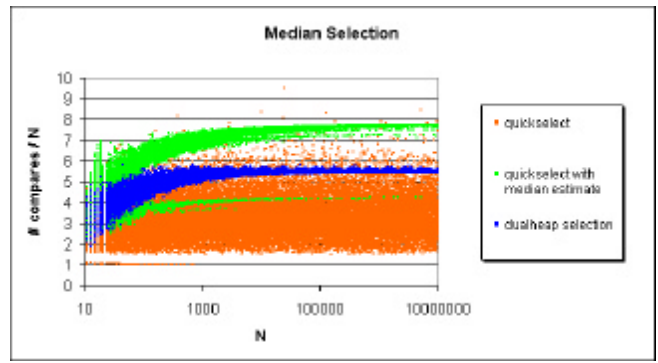

**Figure 3-1. # Compares by Different Selection Algorithms**

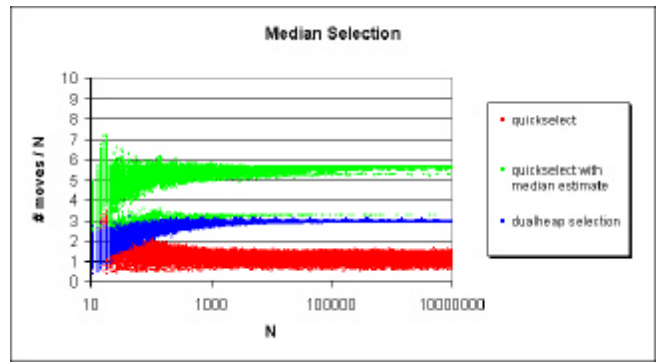

**Figure 3-2. # Moves by Different Selection Algorithms**

## 4. REVIEW

A selection algorithm has been presented that employs a new dualheap data structure and minor variations of R.W. Floyd's 1964 algorithm for bottom-up heap construction. A side effect of the dualheap selection algorithm is that it partitions a set into small and large valued subsets, which is relevant to a wide variety of divide-and-conquer algorithms, including some sorting algorithms.

Empirical tests (complete source code and data available at http://www.eduneer.com/) suggest that the dualheap selection algorithm's serial implementation is competitive with the popular quickselect algorithm. Although the empirical test results seem promising, the dualheap selection algorithm's future depends upon whether its apparently linear worst-case performance can be formally proven.

## 5. CALL FOR ANALYSIS

To encourage dualheap selection algorithm analysis, I'll document at http://www.eduneer.com/ my analysis attempts to prove the worst-case performance of the algorithm's swapping phase is **O**(n). Note that the proof can be limited to the swapping phase since the construction phase's **O**(n) worst-case performance is well known [7].